\documentclass[12pt]{iopart}

\begin{document}

\title[Tao $et$  $al$., Thermoelectric effect of La$T$AsO]{Comparative study on the thermoelectric effect of parent oxypnictides La$T$AsO ($T$ = Fe, Ni)}

\author{Qian Tao$^{1}$, Zengwei Zhu$^{1}$, Xiao Lin$^{1}$, Guanghan Cao$^{1}$, Zhu-an Xu$^{1}$\footnote[3]{Corresponding
author. Tel: (86)571-87953255, E-mail address: zhuan@zju.edu.cn},
Genfu Chen$^{2}$, Jianlin Luo$^{2}$, Nanlin Wang$^{2}$}

\address{$^{1}$ Department of Physics, Zhejiang Unviersity, Hangzhou 310027, China}
\address{$^{2}$ Beijing National Laboratory for Condensed Matter Physics, Institute of Physics, Chinese Academy of Sciences, Beijing 100080, China}

\begin{abstract}
The thermopower and Nernst effect were investigated for undoped
parent compounds LaFeAsO and LaNiAsO. Both thermopower and Nernst
signal in iron-based LaFeAsO are significantly larger than those
in nickel-based LaNiAsO. Furthermore, abrupt changes in both
thermopower and Nernst effect are observed below the structural
phase transition temperature and spin-density wave (SDW) type
antiferromagnetic (AFM) order temperature in Fe-based LaFeAsO. On
the other hand, Nernst effect is very small in the Ni-based
LaNiAsO and it is weakly temperature-dependent, reminiscent of the
case in normal metals. We suggest that the effect of SDW order on
the spin scattering rate should play an important role in the
anomalous temperature dependence of Hall effect and Nernst effect
in LaFeAsO. The contrast behavior between the LaFeAsO and LaNiAsO
systems implies that the LaFeAsO system is fundamentally different
from the LaNiAsO system and this may provide clues to the
mechanism of high $T_c$ superconductivity in the Fe-based systems.

\end{abstract}

\maketitle

\section{\label{sec:level1}Introduction}
The recent discovery of superconductivity in layered 1111 phase
quaternary compounds $R$$T$$Pn$O ($R$ = lanthinides, $T$ = Fe and
Ni, $Pn$ = P and As) has attracted tremendous attention to this
class of materials
\cite{Hosono-LaP,Hosono-LaF,ChenXH-SmOF,WangNL-CeOF,ZhaoZX-LnOD,WenHH-LaSr,WangC-GdTh,LuoJL-LaNi}.
Besides this 1111 phase oxy-pnictides, superconductivity was
subsequently discovered in other iron(nickel)-based layered
compounds with similar Fe(Ni)As layers
\cite{Johrendt-BaK,LiFeAs,WuMK-FeSe,Ogino42622}. In all the
FeAs-based parent compounds, there is a structural phase
transition in the temperature range 100-200 K, and a spin-density
wave (SDW) type antiferromagnetic (AFM) ordering associated with
Fe ions accompanies the structural transition
\cite{WangNL-SDW,DaiPC-LaNeutron,BaoW-BaNeutron}. Various chemical
doping approaches or application of high pressure can suppress the
structural transition and AFM order, and high-$T_c$
superconductivity consequently appears. For example, in the
FeAs-based La-1111 system, superconductivity has been achieved by
the chemical doping at four different crystallographic sites
\cite{Hosono-LaF,WenHH-LaSr,WangC-GdTh,Co1,Co2,LaP}. Meanwhile,
low-$T_c$ superconductivity has been observed in FeP-based
\cite{Hosono-LaP} and NiAs(P)-based \cite{LuoJL-LaNi,Hosono-NiAs}
compounds with similar layered structure, but there is neither
structural transition nor AFM ordering associated with Fe(Ni) ions
in these compounds. Furthermore, many reports have found that both
normal state properties and superconductivity of the NiAs-based
systems are likely of conventional type
\cite{LuoJL-LaNi,Hosono-NiAs}. This result suggests that there
could be a close relationship between structural transition/AFM
order and high-$T_c$ superconductivity. Moreover, the origin of
the AFM order in the parent FeAs-based pnictides is still an open
issue theoretically. Regardless of the origin of the AFM ordering,
several theories have suggested that the superconductivity is tied
to the magnetism in the FeAs-based
materials\cite{Mazin,Yildirim,XiangT,Seo2009}. All these results
imply that the mechanism of superconductivity in FeAs-based
systems could be fundamentally different to NiAs-based systems.
Systematic investigation on the physical properties of these
parent pnictides can shed light on the mechanisms of
superconductivity.

Thermoelectric effects are very sensitive to subtle changes in
electronic structure, and can provide information on the ground
state and low energy excitations. Especially, the transverse
magneto-thermoelectric effect, i.e., Nernst effect, which is
defined as the appearance of a transverse electric field $E_{y}$
in response to a temperature gradient $\nabla T || x$ in the
presence of a perpendicular magnetic field $H||z$ and under open
circuit conditions, has becomes a powerful probe in studying
exotic superconductors like high-$T_c$ cuprates \cite{zhan
nernst}, charge-density wave (CDW) superconductor NbSe$_2$
\cite{NbSe2}, heavy fermion superconductors \cite{heavy fermion},
and $p$-wave superconductor Sr$_2$RuO$_4$ \cite{SrRuO} etc. The
first report on the Nernst effect in F-doped LaFeAsO
superconductor has discovered that the vortex liquid regime below
$T_c$ is quite large \cite{ZhuZW}, consistent with the simulation
result\cite{JPLv}, and there is an enhanced anomalous Nernst
signal just above $T_c$.

Here we report the systematic investigation on the thermopower and
Nernst effect of FeAs-based and NiAs-based parent oxypnictides.
Both thermopower and Nernst coefficient of the undoped LaFeAsO
system are significantly large compared to usual metals and the
strcutural/AFM transition causes anomalous changes in
thermoelectric properties. However, the thermopower and Nernst
effect of the NiAs-based LaNiAsO system are likely of conventional
type, implying usual Fermi liquid behavior. The fundamental
difference in the thermoelectric properties of the two systems
suggests that there is a close relationship between high-$T_c$
superconductivity and anomalous thermoelectric properties in
FeAs-based systems.

\section{\label{sec:level2}Experimental}
The polycrystalline samples of LaFeAsO and LaNiAsO were prepared
by the solid state reaction using LaAs, Fe$_{2}$O$_{3}$/NiO, Fe/Ni
and LaF$_{3}$ as starting materials. The sample preparation
details can be found in the previous report \cite{LuoJL-LaNi}. The
powder X-ray diffraction patterns indicate that the resultant is
single phase and all the diffraction peaks can be well indexed
based on the tetragonal ZrCuSiAs-type structure with the space
group P4/nmm.

The resistivity was measured by usual four-probe method. The Hall
effect was measured by scanning magnetic field at fixed
temperatures. The thermoelectric properties were measured by a
steady-state technique. The temperature gradient used for the
thermoelectric measurements, measured by a pair of differential
Type E thermocouples, was around $0.5$ K/mm. All the measurements
were performed in a Quantum Design PPMS-$9$ system. The Nernst
signal $e_{y}$ is defined as $e_{y}\equiv\frac{E_{y}}{|\nabla T|}$
. The Nernst signal was measured at positive and negative field
polarities, and the difference of the two polarities was taken to
remove any thermopower contributions. The Nernst coefficient
$\nu_N$ is equal to $e_{y}$/$B$. At very low temperatures, $e_{y}$
is not strictly linear with magnetic field $H$, the Nernst
coefficient is then taken as the initial slope of the $e_{y}$
versus $\mu_0H$ curves.

\section{\label{sec:level3}Results and discussion}
\subsection{\label{sec:level1} Thermoelectric effects of LaFeAsO}

Traces of Nernst signal as a function of magnetic field up to
$\mu_0$$H$ of 8 T for the parent oxypnictide LaFeAsO is displayed
in Figure 1 at various temperatures. At high temperatures (see the
lower panel of Fig.1) , the Nernst signal ($e_{y}$) is positive
and changes linearly with magnetic field. The Nernst signal
reaches a maximum around 100 K. Below 100 K, the Nernst signal
decreases, and becomes a little nonlinear with magnetic field as
$T <$ 80 K. Between 40 to 60 K, the Nernst signal changes from
positive to negative. Such a sign change may not suggest the
change in the charge carrier type. In multi-band systems, the
contributions of different bands to the Nernst signal could have
different signs.

The temperature dependence of Nernst coefficient, $\nu_N(T)$,
together with its thermopower, $S(T)$, is shown in Figure 2. The
temperature dependence of thermopower is consistent with previous
reports \cite{Mandrus,LiLJ,LiyYK}, which exhibits a pronounced
hump as the system undergoes the structural phase transition/AFM
ordering at $T^*$ of about 160 K, where $T^*$ is the structural
phase transition temperature and it can be determined by the
resistivity measurements. The Nernst coefficient is defined as the
initial slope of $e_{y}$-$H$ curves for $T<80 K$. The Nernst
coefficient is positive and shows weak temperature dependence as
$T > T^*$ of about 160 K. However $\nu_N$ starts to increase below
$T^*$ of 160 K, reaches a maximum around 100 K, and then decreases
with decreasing temperatures. With further cooling, it changes
sign around 50 K, and then reaches a minimum (valley) around 20 K
before it finally goes to zero. Compared to the usual metals and
the normal state of high-$T_c$ cuprates, $\nu_N(T)$ is much
larger.

In order to get more insight on the thermoelectric effect, the
ratio of $\nu_N$/$S$ is calculated and plotted in the upper panel
of Fig.3. There is a very clear sharp drop in $\nu_N$/$S$ as the
system becomes AFM ordered below $T^*$. But the ratio increases
monotonously with further cooling. By measuring the Hall effect
and thermopower simultaneously, the off-diagonal thermoelectric
(Peltier conductivity) term can be separated from the Nernst
signal and then both the Hall angle tan$\theta$ (defined as
$\rho_{xy}$/$\rho$) and "thermal" Hall angle tan$\theta_{th}$
(defined as $\alpha_{xy}$/$\alpha$) can be obtained. The
temperature dependence of Hall coefficient ($R_{H}$) measured
under magnetic field ($\mu_0H$) of 8 T is also shown in the upper
panel of Fig. 3. $R_H$ is negative, suggesting that the dominant
charge carrier is electron-like, consistent with previous reports
\cite{4NLWang,Wen,23Oak}. The $R_H$ is about -4.8$\times 10^{-9}$
m$^3$/C at $T$ = 300 K, corresponding to a Hall number ($n_H$ =
1/($eR_H$)) of about 0.18 electrons per unit cell, which could be
an upper limit for the electron concentration $n_e$ because
LaFeAsO is a nearly compensated metal according to the band
calculations \cite{BandCal}. For comparison, the more metallic
BaFe$_2$As$_2$ has a Hall number of about 0.56 electrons per unit
cell at room temperature according to the Hall effect
measurements, but band calculations predict a Hall number of 0.15
electrons per unit cell \cite{Wen122}.

The normal state Nernst signal is comprised of two terms, viz.
\begin{equation}
e_{y}=\rho\alpha_{xy}-S\tan\theta=S(\tan\theta_{th}-\tan\theta),
\end{equation}
where $\alpha_{xy}$ is the off-diagonal Peltier coefficient, $S =
\frac{\alpha}{\sigma}$ the thermopower ( $\sigma$ the diagonal
conductivity and $\alpha$ the diagonal Peltier coefficient),
$\rho$ the resisitivity, tan$\theta$ the Hall angle, and
tan$\theta_{th}$ the "thermal" Hall angle. For usual simple
metals, the two terms related to the Hall angle and "thermal" Hall
angle in Eq. (1) cancel each other, which is so-called "Sondheimer
cancellation" \cite{Cancellation,Cancellation1}. It is known that
Hall angle, rather than Hall coefficient, is directly related to
the scattering rate of the quasiparticle scattering. The Hall
angle can be calculated from the Hall coefficient, i.e.,
tan$\theta$ = $\rho_{xy}$/$\rho$, and then the "thermal" Hall
angle can be obtained from the Nernst signal by using Eq.(1). The
obtained Hall angle and "thermal" Hall angle are plotted in the
lower panel of Fig.3 as a function of temperature. The two angles
exhibit very different temperature dependence, especially the
decrease in the "thermal" Hall angle is much more significant
below $T^*$. It becomes clear that the Sondheimer cancellation of
the Nernst signal is no longer held for LaFeAsO because of the
different temperature dependence of the two angles below $T^*$.
Even above $T^*$, the Nernst coefficient is also larger than that
of usual metals. The enhanced Nernst signal above $T^*$ could
result from the multi-band effect. The presence of two types of
charge carriers in multi-band systems such as
NbSe$_2$\cite{NbSe2}, could invalidate the Sondheimer
cancellation, resulting in a large Nernst signal. However it is
hard to understand the anomalous temperature dependence of $\nu_N$
below $T^*$ even in the frame of multi-band effect. There might be
significant changes in the scattering mechanism below $T^*$, and
these changes seem to have more pronounced influence on the
thermal channel. We propose that the SDW order or SDW fluctuations
could affect the spin-dependent scattering process and thus cause
significant changes in the scattering rates which could be
band-dependent. Subtle changes in the scattering mechanism could
causes anomalous Nernst effect, as observed in the p-wave
superconductor Sr$_2$RuO$_4$\cite{SrRuO}.

The first-principles band calculations have proposed that the
nesting between the electron-type Fermi surface (FS) and hole-type
FS could account for the AFM transition in the parent compounds
such as LaFeAsO, and that the superconductive pairing might be
mediated by spin fluctuations \cite{WangNL-SDW,Mazin,BandCal}.
Moreover, the studies of angle-resolved photoemission spectroscopy
(ARPES) discovered that the coexistence of hole and electron
pockets connected via the AFM wave vector is essential to
high-$T_c$ superconductivity \cite{ARPES1,ARPES2}. The AFM order
and structural phase transition in the parent compounds causes
significant changes in the electronic structure. Based on our
measurements of resistivity and Hall effect, the charge carrier
concentration in LaFeAsO estimated by the Hall number decreases by
a factor of about 100 and meanwhile the scattering rate (in
inverse proportion to the mobility) decreases by a factor of about
160 as temperature decreases from 300 K to 10 K. For the parent
compound BaFe$_2$As$_2$, similar results were reported
\cite{Wen122,Hall}. Such changes in the charge carrier
concentration and scattering rates in the parent compound
BaFe$_2$As$_2$, SrFe$_2$As$_2$, and EuFe$_2$As$_2$ were also
reported by the optical spectroscopy measurements \cite{Wang,
Eu122}. All the results suggest that the charge carrier transport
are dominated by the AFM fluctuations in the parent compounds.
When the system undergoes the AFM order below about 160 K, the
charge carrier concentration deceases sharply due to the SDW
gapping on FSs, but the scattering rates decreases even more
drastically due to the AFM order which suppresses the spin
fluctuations. It can be seen from Fig.3, the AFM order may have
more pronounced influence on the thermal channel. Thus our results
imply that the anomalous changes in the thermopower and Nernst
coefficient in LaFeAsO should have close relation with the
inter-band scattering between electron-type and hole-type bands.

\subsection{\label{sec:level2} Thermoelectric effects of LaNiAsO}
The thermopower and Hall coefficient (measured under $\mu_0H$ of 5
T) of LaNiAsO are plotted as a function of temperature in Fig.4.
The negative $R_H$ implies that the charge carriers are dominantly
electron type, same as in LaFeAsO. The absolute value of $R_H$ for
LaNiAsO is more than 1 order of magnitude smaller than that of
LaFeAsO, possibly indicating that the LaNiAsO system has a
relatively higher carrier density. The Hall number $n_H$ at $T$ =
300 K is about 8.3 electrons per unit cell, about 50 times larger
compared to LaFeAsO. Since Ni$^{2+}$3$d^8$ contributes two more
electrons than Fe ion does, the Fermi energy shifts up in LaNiAsO,
and the hole bands tend to be fully filled. As a result predicted
by band calculations\cite{BandCal}, the electron bands dominate
the conductivity in LaNiAsO \cite{LuoJL-LaNi}. The much smaller
thermopower also suggests that LaNiAsO is a good metal with high
density of charge carriers compared to LaFeAsO. The small hollow
in $S$ at low temperature could be caused by phonon drag effect.
It is interesting that $R_H$ also exhibits a sharp decrease below
50 K. Such a change in $R_H$ at low temperature has not been well
understood presently.

Traces of Nernst signal as a function of magnetic field for
LaNiAsO is displayed in Fig.5 at several selected temperatures .
The Nernst signal ($e_{y}$) is very small, comparable to the noise
level (the noise voltage is about 10 nV in our measurement
system). This Nernst signal is as small as usual metals. $e_{y}$
at $\mu_0H$ of 6 T is only in the range of $\pm$30 nV/K, almost
two order of magnitude smaller than that of LaFeAsO. The Nernst
coefficient is obtained by fitting the $e_y$ vs. $H$ curves with a
linear function, and it is plotted in Fig.6. The Hall angle
tan$\theta$ and the term $S$tan$\theta$ are also plotted in Fig.6
for comparison. It can seen that the Nernst coefficient is
comparable to the term $S$tan$\theta$, implying that the
Sondheimer cancellation is partially held and the system could be
dominated by only one type charge carrier (electron). Please note
that the temperature dependence of tan$\theta$ is very weak and
its magnitude is much smaller that that of LaFeAsO, especially at
low temperatures.

The result of Nernst effect suggests that LaNiAsO is like usual
metal and only one type (electron-type) charge carrier dominates.
Xu et al. \cite {BandCal} compared the band structures of LaFeAsO
and LaNiAsO by first-princeples calculations. It was found that
the electron FS cylinders around M point becomes larger for
LaNiAsO, and the hole-type FS cylinders around $\Gamma$ point
disappear. Thus the nesting between hole-type FSs and
electron-type FSs proposed in LaFeAsO is no longer held in
LaNiAsO. Recall that the multi-band effect could account for the
enhanced Nernst signals in both parent and F-doped LaFeAsO which
has relatively high superconducitng transition temperatures, the
inter-band scattering should indeed play an important role in the
occurrence of high $T_c$. On the other hand, our result suggests
that the LaNiAsO system lacks of such an inter-band scattering
mechanism, and thus it has a low $T_c$. The relationship between
high-$T_c$ superconductivity and the inter-band scattering is
worthy of further experimental investigations.

\section{\label{sec:level4}Conclusions}

In summary, we report the thermeropower and Nernst effect of
FeAs-based and NiAs-based parent oxypnictides. For the LaFeAsO
system, it is found that both thermopower and Nernst coefficient
are significantly large compared to usual metals and the
structural/AFM transition associated with Fe ions causes
significant enhancements in the thermoelectric coefficients. We
propose that the unique strong inter-band scattering as well as
electron correlation in FeAs-based systems may account for these
anomalies in thermoelectric properties. Meanwhile, the thermopower
and Nernst effect of the NiAs-based LaNiAsO system are of
conventional type, implying usual Fermi liquid behavior. The
fundamental difference in the thermoelectric properties of the two
systems suggests that there is a close relationship between
high-$Tc$ superconductivity and anomalous thermoelectric
properties in FeAs-based pnictides.

\section*{Acknowledgments}
This project is supported by the National Science Foundation of
China (Grant. No.10628408 and 10931160425) and the National Basic
Research Program of China (Grant No. 2006CB601003 and
2007CB925001).

\pagebreak[4]

\begin{figure}
\caption{\label{Fig1}(Color Online) Magnetic field dependence of
the Nernst signal ($e_{y}$) at different temperatures for
LaFeAsO.}
\end{figure}

\begin{figure}
\caption{\label{Fig2} (Color Online) Temperature dependence of the
Nernst coefficient for LaFeAsO. The thermopower measured under
zero magnetic field is shown together. }
\end{figure}

\begin{figure}
\caption{\label{Fig3} Upper Panel: The Hall coefficient ($R_{H}$)
and the ratio of $\nu_N$/$S$ versus temperature for LaFeAsO. The
Hall effect was measured under magnetic field ($\mu_0H$) of 8 T.
Lower Panel: Temperature dependence of Hall angle tan$\theta$ and
"thermal" Hall angle tan$\theta_{th}$ at $\mu_0H$ of 8 T for
LaFeAsO. See text for more details.}
\end{figure}

\begin{figure}
\caption{\label{Fig4} (Color Online) Temperature dependence of the
thermopower and Hall coefficient for LaNiAsO. The Hall effect was
measured under magnetic field ($\mu_0H$) of 5 T.}
\end{figure}

\begin{figure}
\caption{\label{Fig5} (Color Online) Magnetic field dependence of
the Nernst signal ($e_{y}$) at selected temperatures for LaNiAsO.
}
\end{figure}

\begin{figure}
\caption{\label{Fig6} (Color Online) Nernst coefficient versus
temperature for LaNiAsO. The Hall angle tan$\theta$ and the term
$S$tan$\theta$ are also plotted, where the Hall angle tan$\theta$
is normalized to $\mu_0H$ of 1 T.}
\end{figure}

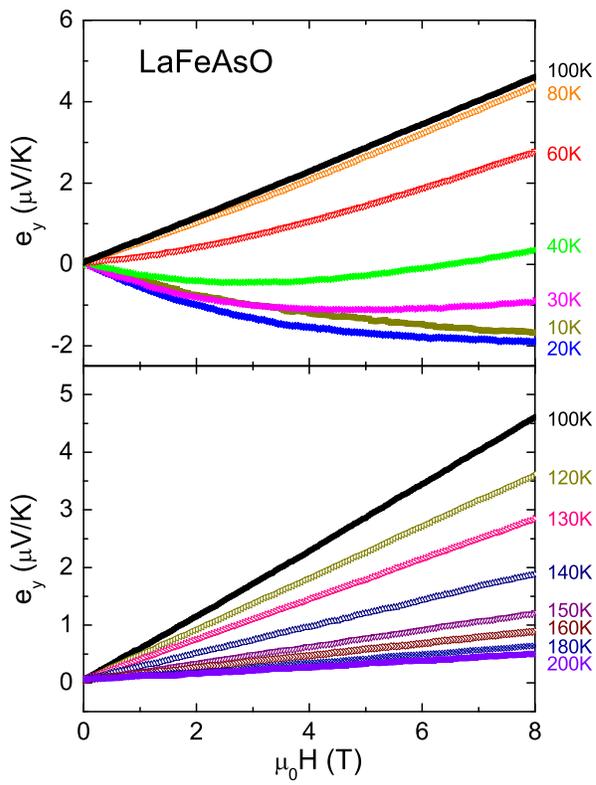
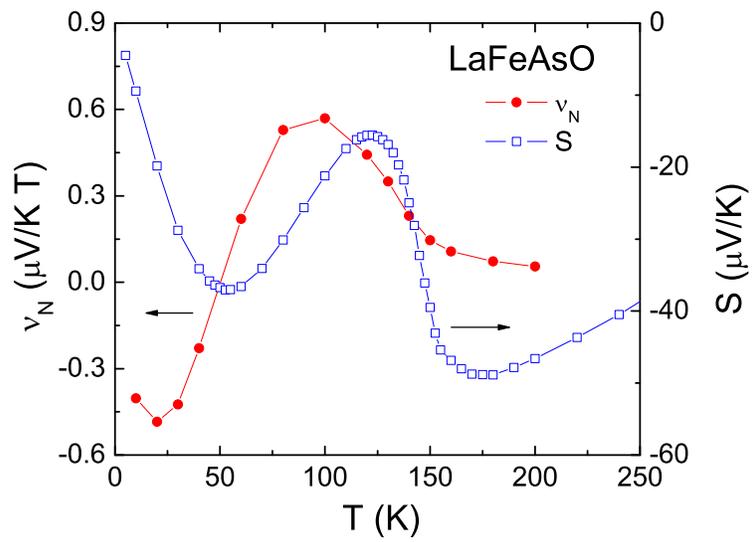
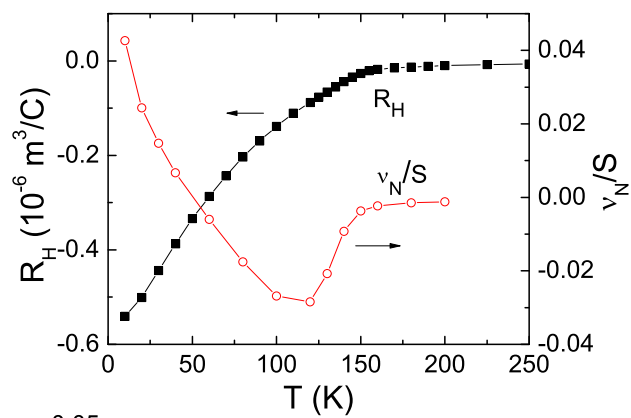
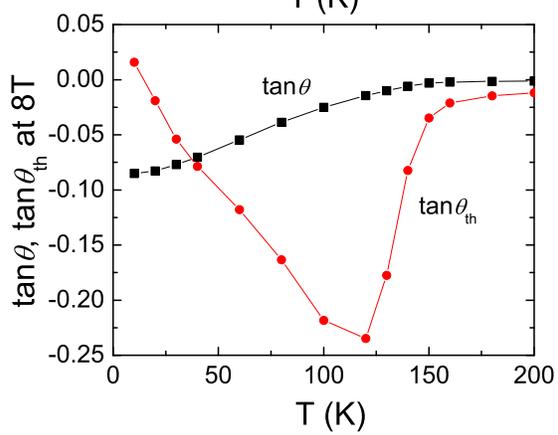
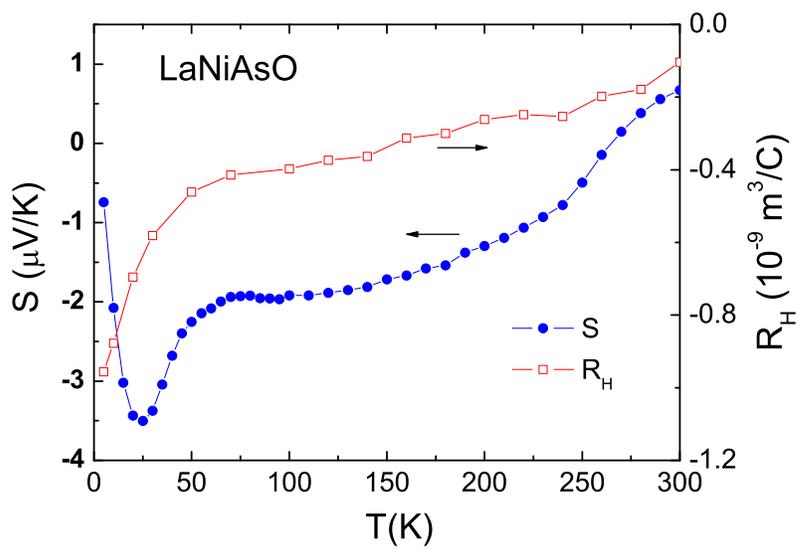
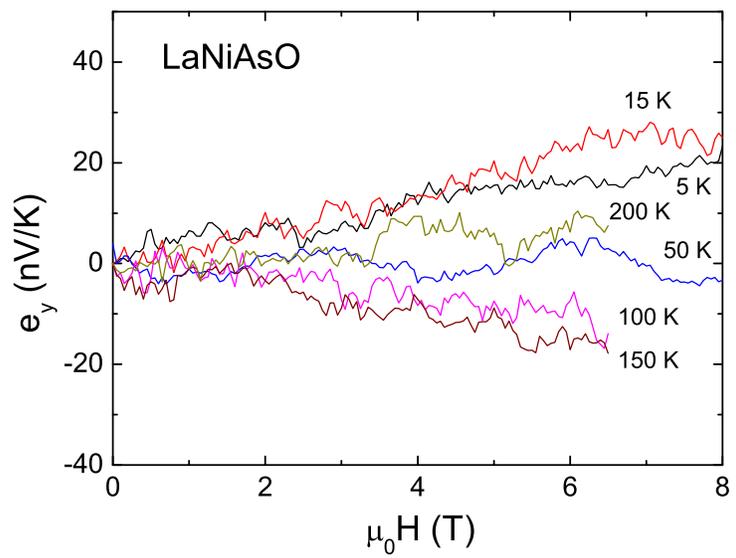
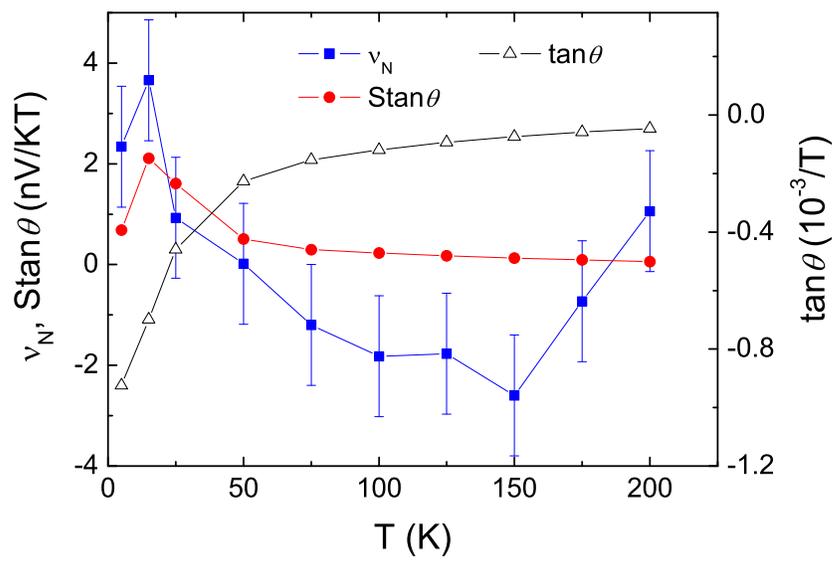
\end{document}